\documentclass[aps,prb,twocolumn,superscriptaddress,floatfix,amsmath]{revtex4}

\usepackage{graphicx}
\usepackage{amsmath}
\usepackage{amssymb}

\renewcommand{\vec}[1]{\boldsymbol{#1}}

\begin{document}

\title{Doping and temperature dependence of electron spectrum and
quasiparticle dispersion in doped bilayer cuprates}

\author{Yu Lan}
\affiliation{Department of Physics, Beijing Normal University,
Beijing 100875, China}

\author{Jihong Qin}
\affiliation{Department of Physics, Beijing University of Science
and Technology, Beijing 100083, China}

\author{Shiping Feng}
\affiliation{Department of Physics, Beijing Normal University,
Beijing 100875, China}


\begin{abstract}
Within the $t$-$t'$-$J$ model, the electron spectrum and
quasiparticle dispersion in doped bilayer cuprates in the normal
state are discussed by considering the bilayer interaction. It is
shown that the bilayer interaction splits the electron spectrum of
doped bilayer cuprates into the bonding and antibonding components
around the $[\pi,0]$ point. The differentiation between the bonding
and antibonding components is essential, which leads to two main
flat bands around the $[\pi,0]$ point below the Fermi energy. In
analogy to the doped single layer cuprates, the lowest energy states
in doped bilayer cuprates are located at the $[\pi/2,\pi/2]$ point.
Our results also show that the striking behavior of the electronic
structure in doped bilayer cuprates is intriguingly related to the
bilayer interaction together with strong coupling between the
electron quasiparticles and collective magnetic excitations.

\end{abstract}


\maketitle


\section{Introduction}

Cuprate materials are unusual in that the undoped cuprates are
insulators with an antiferromagnetic (AF) long-range-order (AFLRO),
and changing the carrier concentration by ionic substitution or
increase of the oxygen content turns these compounds into correlated
metals leaving short-range AF correlations still intact
\cite{bedell,kastner,shen,campuzano,shen1}. The single common
feature of cuprate superconductors is the presence of the
two-dimensional CuO$_{2}$ plane
\cite{bedell,kastner,shen,campuzano,shen1}, and it seems evident
that the unusual behaviors of doped cuprates are dominated by the
CuO$_{2}$ plane \cite{anderson}. This layered crystal structure
leads to that cuprate superconductors are highly anisotropic
materials, then the electron spectral function $A({\bf k},\omega)$
is dependent on the in-plane momentum
\cite{bedell,kastner,shen,campuzano,shen1}. After twenty years
extensive studies, it has been shown that many of the unusual
physical properties have often been attributed to particular
characteristics of low energy excitations determined by the
electronic structure \cite{shen,campuzano,shen1}.

Although the electronic structure of doped cuprates is
well-established by now \cite{shen,campuzano,shen1}, its full
understanding is still a challenging issue. Experimentally,
angle-resolved photoemission spectroscopy (ARPES) experiments have
provided rather detailed information on the electronic structure of
the doped single layer and bilayer cuprates
\cite{shen,campuzano,shen1,shen2,dessau,dfeng,kordyuk,chuang,campuzano1}.
An important issue is whether the behavior of the low energy
excitations determined by the electronic structure is universal or
not. The early ARPES measurements \cite{shen1,shen2,dessau} showed
that the charge carriers doped into the parent compounds first enter
into the ${\bf k}= [\pi/2,\pi/2]$ (in units of inverse lattice
constant) point in the Brillouin zone. Moreover, the electron
spectral function $A({\bf k},\omega)$ has a flat band form as a
function of energy $\omega$ for momentum ${\bf k}$ in the vicinity
of the $[\pi,0]$ point, which leads to the unusual quasiparticle
dispersion around the $[\pi,0]$ point with anomalously small changes
of electron energy as a function of momentum
\cite{shen1,shen2,dessau}. Furthermore, this flat band is just below
the Fermi energy. Recently, the improvements in the resolution of
the ARPES experiments \cite{shen,campuzano,dfeng,kordyuk,chuang}
allowed to resolve additional features in the electron spectral
function $A({\bf k},\omega)$. Among these new achievements is the
observation of the bilayer splitting (BS) in doped bilayer cuprates
in a wide doping range \cite{shen,dfeng,kordyuk,chuang}. In this
case, whether the electronic structure of doped cuprates can be
influenced by the interaction between CuO$_{2}$ planes has been an
interesting issue. The study of the electronic structure is
complicated by the BS, that is, the BS of the CuO$_{2}$ plane
derived the electronic structure in the bonding and antibonding
bands due to the present of CuO$_{2}$ bilayer blocks in the unit
cell of doped bilayer cuprates \cite{dfeng,kordyuk,chuang}. The
magnitude of the BS is the doping independent, and increases upon
approaching the $[\pi,0]$ point, where the BS exhibits the largest
value. As a result of the maximal BS at the $[\pi,0]$ point, there
are two main flat bands around the $[\pi,0]$ point
\cite{dfeng,kordyuk,chuang}. Moreover, it has been shown that a well
pronounced peak-dip-hump structure in the electron spectrum of doped
bilayer cuprates in the superconducting state is partially caused by
the BS \cite{kordyuk1,borisenko}. Theoretically, the
electron-removal spectral functions at the $[\pi,0]$ point and
quasiparticle dispersion of the multilayer cuprates in the normal
state has been discussed \cite{mori} within the two-dimensional
multilayer $t$-$t'$-$t''$-$J$ model based on the resonating valence
bond wave function and Gutzwiller approximation, in particular,
their results show that there are two main flat bands around the
$[\pi,0]$ point in the bilayer system. However, the resonating
valence bond wave function and Gutzwiller approximation can be only
applied to discuss the zero temperature physical properties of doped
cuprates. To the best of our knowledge, the doping and temperature
dependence of the electron spectrum and quasiparticle dispersion in
both doped single layer and bilayer cuprates have not been treated
from a unified point of view for the normal state.

In our earlier work using the charge-spin separation (CSS)
fermion-spin theory \cite{feng}, the electronic structure of the
doped single layer cuprates has been calculated within the single
layer $t$-$t'$-$J$ model \cite{guo}, and the obtained doping
dependence of the electron spectrum and quasiparticle dispersion are
consistent with the corresponding ARPES experiments
\cite{shen1,shen2,dessau}. In this paper we show explicitly if the
bilayer interaction is included, one can reproduce some main
features in the normal state observed experimentally on doped
bilayer cuprates \cite{shen,dfeng,kordyuk,chuang}. The
differentiation between the bonding and antibonding components is
essential. Two main flat bands around the $[\pi,0]$ point are rather
similar to the scenario argued in Refs. \onlinecite{dfeng},
\cite{kordyuk}, and \cite{mori} for the normal state, and the
quasiparticle dispersion we derive from the bilayer $t$-$t'$-$J$
model (without additional terms and adjustable parameters)
demonstrates explicitly this energy band splitting. In comparison
with the case of the doped single layer cuprates \cite{guo}, our
present results also show that the striking behavior of the
electronic structure in doped bilayer cuprates is intriguingly
related to the bilayer interaction together with strong coupling
between the electron quasiparticles and collective magnetic
excitations.

The rest of this paper is organized as follows. In Sec. II, we
introduce the bilayer $t$-$t'$-$J$ model that include the bilayer
hopping and bilayer magnetic exchange interaction. Within this
bilayer $t$-$t'$-$J$ model, we calculate explicitly the longitudinal
and transverse components of the electron Green's functions based on
the CSS fermion-spin theory by considering charge carrier
fluctuations, and then obtain the bonding and antibonding electron
spectral functions according to these longitudinal and transverse
components of the electron Green's functions. The doping and
temperature dependence of the electron spectrum and quasiparticle
dispersion of doped bilayer cuprates in the normal state are
presented in Sec. III. Sec. IV is devoted to a summary and
discussions.

\section{Theoretical Framework}

The basic element of cuprate materials is two-dimensional CuO$_{2}$
plane, and it is believed that the unusual behaviors of doped
cuprates are closely related to the doped CuO$_{2}$ plane as
mentioned above \cite{bedell,kastner,shen,campuzano,shen1}. It has
been argued that the essential physics of the doped CuO$_{2}$ plane
is contained in the $t$-$t'$-$J$ model on a square lattice
\cite{anderson}. In this case, the $t$-$t'$-$J$ model in the bilayer
structure is expressed as,
\begin{eqnarray}
H&=&-t\sum_{i\hat{\eta}a\sigma}C^{\dagger}_{ia\sigma}
C_{i+\hat{\eta}a\sigma}+t'\sum_{i\hat{\tau}a\sigma}
C^{\dagger}_{ia\sigma}C_{i+\hat{\tau}a\sigma}\nonumber\\
&-&\sum_{i\sigma}
t_{\perp}(i)(C^{\dagger}_{i1\sigma}C_{i2\sigma}+H.c.)\
+\mu\sum_{ia\sigma}C^{\dagger}_{ia\sigma}C_{ia\sigma}\nonumber\\
&+&J\sum_{i\hat{\eta}a}{\bf S}_{ia} \cdot {\bf S}_{i+\hat{\eta}a}
+J_{\perp}\sum_{i}{\bf S}_{i1} \cdot {\bf S}_{i2},
\end{eqnarray}
where $\hat{\eta}=\pm\hat{x},\pm\hat{y}$, $\hat{\tau}=\pm\hat{x}
\pm\hat{y}$, $a=1,2$ is plane index, $C^{\dagger}_{ia\sigma}$
($C_{ia\sigma}$) is the electron creation (annihilation) operator,
${\bf S}_{ia}= C^{\dagger}_{ia}{\vec\sigma} C_{ia}/2$ is the spin
operator with ${\vec\sigma}=(\sigma_{x},\sigma_{y},\sigma_{z})$ as
Pauli matrices, $\mu$ is the chemical potential, and the interlayer
hopping \cite{chakarvarty},
\begin{eqnarray}
t_{\perp}({\bf k})={t_{\perp}\over 4}(\cos k_{x} -\cos k_{y})^{2},
\end{eqnarray}
describes coherent hopping between the CuO$_{2}$ planes. This
functional form of the interlayer hopping in Eq. (2) is predicted on
the basis of the local density approximation calculations
\cite{chakarvarty}, and later the experimental observed BS agrees
well with it \cite{shen,dfeng,kordyuk,chuang}. The $t$-$t'$-$J$
Hamiltonian (1) is supplemented by the single occupancy local
constraint $\sum_\sigma C_{ia\sigma}^{\dagger}C_{ia\sigma}\leq 1$.
This local constraint can be treated properly in analytical
calculations within the CSS fermion-spin theory \cite{feng}, where
the constrained electron operators in the $t$-$J$ type model are
decoupled as $C_{ia\uparrow}= h^{\dagger}_{ia\uparrow}S^{-}_{ia}$
and $C_{ia\downarrow}= h^{\dagger}_{ia\downarrow}S^{+}_{ia}$, with
the spinful fermion operator $h_{ia\sigma}= e^{-i\Phi_{ia\sigma}}
h_{ia}$ keeps track of the charge degree of freedom together with
some effects of the spin configuration rearrangements due to the
presence of the doped hole itself (dressed holon), while the spin
operator $S_{ia}$ keeps track of the spin degree of freedom. The
advantage of this CSS fermion-spin theory is that the electron local
constraint for single occupancy,
$\sum_{\sigma}C^{\dagger}_{ia\sigma}
C_{ia\sigma}=S^{+}_{ia}h_{ia\uparrow}h^{\dagger}_{ia\uparrow}
S^{-}_{ia}+S^{-}_{ia}h_{ia\downarrow}h^{\dagger}_{ia\downarrow}
S^{+}_{ia}=h_{ia}h^{\dagger}_{ia}(S^{+}_{ia}S^{-}_{ia}+S^{-}_{ia}
S^{+}_{ia})=1-h^{\dagger}_{ia} h_{ia}\leq 1$, is satisfied in
analytical calculations, and the double spinful fermion occupancies
$h^{\dagger}_{ia\sigma} h^{\dagger}_{ia-\sigma}=
e^{i\Phi_{ia\sigma}}h^{\dagger}_{ia} h^{\dagger}_{ia}
e^{i\Phi_{ia-\sigma}}=0$ and $h_{ia\sigma} h_{ia-\sigma}=
e^{-i\Phi_{ia\sigma}}h_{ia}h_{ia} e^{-i\Phi_{ia-\sigma}}=0$, are
ruled out automatically. It has been shown \cite{feng} that these
dressed holons and spins are gauge invariant, and in this sense they
are real and can be interpreted as the physical excitations
\cite{laughlin}. Although in common sense $h_{ia\sigma}$ is not a
real spinful fermion, it behaves like a spinful fermion. In this CSS
fermion-spin representation, the low-energy behavior of the bilayer
$t$-$t'$-$J$ Hamiltonian (1) can be expressed as,
\begin{eqnarray}
H&=&t\sum_{i\hat{\eta}a}(h^{\dagger}_{i+\hat{\eta}a\uparrow}
h_{ia\uparrow}S^{+}_{ia}S^{-}_{i+\hat{\eta}a}+
h^{\dagger}_{i+\hat{\eta}a\downarrow}h_{ia\downarrow}S^{-}_{ia}
S^{+}_{i+\hat{\eta}a})\nonumber\\
&-&t'\sum_{i\hat{\tau}a} (h^{\dagger}_{i+\hat{\tau}a\uparrow}
h_{ia\uparrow}S^{+}_{ia} S^{-}_{i+\hat{\tau}a}+
h^{\dagger}_{i+\hat{\tau}a\downarrow}h_{ia\downarrow}
S^{-}_{ia}S^{+}_{i+\hat{\tau}a})\nonumber\\
&+&\sum_{i}t_{\perp}(i)
(h^{\dagger}_{i2\uparrow}h_{i1\uparrow}S^{+}_{i1}S^{-}_{i2}+
h^{\dagger}_{i1\uparrow}h_{i2\uparrow}S^{+}_{i2}S^{-}_{i1}\nonumber\\
&+&h^{\dagger}_{i2\downarrow}h_{i1\downarrow}S^{-}_{i1} S^{+}_{i2}
+h^{\dagger}_{i1\downarrow}h_{i2\downarrow}S^{-}_{i2}
S^{+}_{i1})-\mu\sum_{ia\sigma}h^{\dagger}_{ia\sigma}h_{ia\sigma}\nonumber\\
&+&{J_{\rm eff}}\sum_{i\hat{\eta}a}{\bf S}_{ia}\cdot {\bf
S}_{i+\hat{\eta}a}+ {J_{\rm eff\perp}}\sum_{i}{\bf S}_{i1}\cdot {\bf
S}_{i2},
\end{eqnarray}
where $J_{\rm eff}=J(1-\delta)^{2}$, $J_{\rm eff\perp}=J_{\perp}
(1-\delta)^{2}$, and $\delta=\langle h^{\dagger}_{ia\sigma}
h_{ia\sigma}\rangle=\langle h^{\dagger}_{ia} h_{ia}\rangle$ is the
hole doping concentration. As a result, the magnetic energy term in
the bilayer $t$-$t'$-$J$ model is only to form an adequate spin
configuration \cite{anderson1}, while the kinetic energy part has
been expressed as the interaction between the dressed holons and
spins, and then this interaction dominates the essential physics of
doped cuprates \cite{feng}.

Within the CSS fermion-spin theory, the electron spectral function
and related quasiparticle dispersion of the doped single layer
cuprates in the normal state have been discussed by considering the
dressed holon fluctuations around the mean-field (MF) solution
\cite{guo}, where the spin part is limited to the first-order (the
MF level), while the full dressed holon Green's function is treated
self-consistently in terms of the Eliashberg's strong coupling
theory \cite{eliashberg}. For doped bilayer cuprates, there are two
coupled CuO$_{2}$ planes in the unit cell. This leads to that the
energy spectrum has two branches, then the dressed holon and spin
Green's functions are matrices, and can be expressed as $g({\bf k},
\omega)=g_{L}({\bf k},\omega)+\sigma_{x} g_{T}({\bf k}, \omega)$ and
$D({\bf k},\omega)=D_{L}({\bf k}, \omega)+\sigma_{x} D_{T}({\bf k},
\omega)$, respectively, where the longitudinal and transverse parts
are corresponding to the in-layer and inter-layer Green's functions
\cite{feng1}. In this case, we follow the previous discussions for
the single layer case \cite{guo}, and obtain explicitly the
longitudinal and transverse parts of the full dressed holon Green's
functions of the doped bilayer system as [see the Appendix],
\begin{subequations}
\begin{eqnarray}
g_{L}({\bf k},\omega)&=&{1\over 2}\sum_{\nu=1,2}{Z^{(\nu)}_{FA}
\over\omega-\bar{\xi}_{\nu{\bf k}}}, \\
g_{T}({\bf k},\omega)&=&{1\over 2}\sum_{\nu=1,2}(-1)^{\nu+1}{
Z^{(\nu)}_{FA }\over \omega-\bar{\xi}_{\nu{\bf k}}},
\end{eqnarray}
\end{subequations}
with the renormalized dressed holon excitation spectrum
${\bar\xi}_{\nu{\bf k}}=Z^{(\nu)}_{FA}\xi_{\nu{\bf k}}$, where the
MF dressed holon excitation spectrum is given as,
\begin{eqnarray}
\xi_{\nu{\bf k}}=Zt\chi_{1}\gamma_{\bf{k}}-Zt'\chi_{2}
\gamma'_{\bf{k}}-\mu+(-1)^{\nu+1}\chi_{\perp}t_{\perp}({\bf{k}}),
\end{eqnarray}
with the spin correlation functions $\chi_{1}=\langle S_{ia}^{+}
S_{i+\hat{\eta}a}^{-} \rangle$, $\chi_{2}=\langle S_{ia}^{+}
S_{i+\hat{\tau}a}^{-} \rangle$, $\chi_{\perp}=\langle S^{+}_{i1}
S^{-}_{i2}\rangle$, $\gamma_{\bf{k}}=(1/Z)\sum_{\hat{\eta}}
{e^{i{\bf{k}}\cdot{\hat{\eta}}}}$, $\gamma'_{\bf{k}}=(1/Z)
\sum_{\hat{\tau}}{e^{i{\bf{k}} \cdot{\hat{\tau}}}}$, and $Z$ is the
number of the nearest neighbor or second-nearest neighbor sites,
while the quasiparticle coherent weights $Z^{(1)-1}_{FA}=
Z^{-1}_{F1}- Z^{-1}_{F2}$ and $Z^{(2)-1}_{FA}=Z^{-1}_{F1}+
Z^{-1}_{F2}$, with $Z^{-1}_{F1}=1- \Sigma^{(ho)}_{L}({\bf k}_{0},
\omega)\mid_{\omega=0}$, $Z^{-1}_{F2}=\Sigma^{(ho)}_{T}({\bf k}_{0},
\omega) \mid_{\omega=0}$, and ${\bf k}_{0}=[\pi/2,\pi/2]$, where
$\Sigma^{(ho)}_{L}({\bf k},\omega)$ and $\Sigma^{(ho)}_{T}({\bf k},
\omega)$ are the corresponding antisymmetric parts of the
longitudinal and transverse dressed holon self-energy functions,
\begin{subequations}
\begin{eqnarray}
&\Sigma^{(h)}_{L}&({\bf k},i\omega_{n})={1\over N^{2}}\sum_{\bf
p,q}[R^{(1)}_{{\bf{p+q+k}}}{1\over\beta}\sum_{ip_{m}}g_{L}({\bf
p+k},ip_{m}+i\omega_{n})\nonumber\\
&\times&\Pi_{LL}({\bf p},{\bf q},ip_{m})+R^{(2)}_{{\bf p+q+k}}
{1\over\beta}\sum_{ip_{m}}g_{T}({\bf p+k},ip_{m}+i\omega_{n})
\nonumber\\
&\times& \Pi_{TL}({\bf p},{\bf q},ip_{m})],\\
&\Sigma^{(h)}_{T}&({\bf k},i\omega_{n})={1\over N^{2}}\sum_{\bf p,
q}[R^{(1)}_{{\bf{p+q+k}}}{1\over\beta} \sum_{ip_{m}}g_{T}({\bf
p+k},ip_{m}+i\omega_{n})\nonumber \\
&\times&\Pi_{TT}({\bf p},{\bf q},ip_{m})+R^{(2)}_{{\bf p+q+k}}
{1\over\beta}\sum_{ip_{m}}g_{L}({\bf p+k}, ip_{m} +i\omega_{n})
\nonumber\\
&\times&\Pi_{LT}({\bf p},{\bf q},ip_{m})],
\end{eqnarray}
\end{subequations}
with $R^{(1)}_{{\bf k}}=[Z(t\gamma_{\bf k}-t'\gamma'_{\bf k})]^{2}
+t_{\perp}^{2}({\bf k})$, $R^{(2)}_{{\bf k}}=2Z(t\gamma_{\bf k}-t'
\gamma'_{\bf k})t_{\perp}({\bf k})$, and the spin bubbles
$\Pi_{\eta,\eta'}({\bf p},{\bf q},ip_{m})=(1/\beta)\sum_{iq_{m}}
D^{(0)}_{\eta}({\bf q},iq_{m}) D^{(0)}_{\eta'}({\bf q+p},
iq_{m}+ip_{m})$, where $\eta=L,T$, $\eta'=L,T$, and the longitudinal
and transverse parts of the MF spin Green's functions,
\begin{subequations}
\begin{eqnarray}
D^{(0)}_{L}({\bf k},\omega)&=&{1\over 2}\sum_{\nu=1,2}
{B_{\nu{\bf k}}\over \omega^{2}-\omega^{2}_{\nu{\bf k}}}, \\
D^{(0)}_{T}({\bf k},\omega)&=&{1\over 2}
\sum_{\nu=1,2}(-1)^{\nu+1}{B_{\nu{\bf k}}\over \omega^{2}-
\omega^{2}_{\nu{\bf k}}},
\end{eqnarray}
\end{subequations}
with $B_{\nu {\bf k}}=\lambda (2\epsilon_{\parallel}\chi_{1}^{z}
+\chi_{1})\gamma_{\bf{k}}-2\lambda{'}\chi_{2}^{z}\gamma'_{\bf{k}}
-J_{\rm eff \perp}[\chi_{\perp}+2\chi_{\perp}^{z}(-1)^{\nu}]
\epsilon_{\perp}({\bf{k}})-\lambda(\epsilon_{\parallel}\chi_{1}+
2\chi_{1}^{z})+\lambda'\chi_{2}-J_{\rm eff\perp}[2\chi_{\perp}^{z}
+\chi_{\perp}(-1)^{\nu}]$, $\lambda=2ZJ_{\rm eff}$, $\lambda'=4Z
\phi_{2}t'$, $\epsilon_{\parallel}=1+2t\phi_{1}/J_{\rm eff}$,
$\epsilon_{\perp}({\bf{k}})=1+4\phi_{\perp}t_{\perp}({\bf{k}})
/J_{\rm eff\perp}$, the spin correlation functions
$\chi_{1}^{z}=\langle S_{ia}^{z} S_{i+\hat{\eta}a}^{z}\rangle$,
$\chi_{2}^{z}=\langle S_{ia}^{z} S_{i+\hat{\tau}a}^{z}\rangle$,
$\chi^{z}_{\perp}=\langle S_{i1}^{z} S_{i2}^{z}\rangle$, the dressed
holon particle-hole order parameters $\phi_{1}=\langle
h^{\dagger}_{ia\sigma} h_{i+\hat{\eta}a\sigma}\rangle$,
$\phi_{2}=\langle h^{\dagger}_{ia\sigma}
h_{i+\hat{\tau}a\sigma}\rangle$, $\phi_{\perp}=\langle
h^{\dagger}_{i1\sigma}h_{i2\sigma}\rangle$, and the MF dressed spin
excitation spectrum,
\begin{eqnarray}
\omega^{2}_{\nu{\bf k}}&=&\lambda^{2}[(A_{2}-\alpha
\epsilon_{\parallel}\chi_{1}^{z}\gamma_{\bf k}-{1\over 2Z}\alpha
\epsilon_{\parallel}\chi_{1})(1-\epsilon_{\parallel}\gamma_{\bf k})
\nonumber\\
&+&{1 \over 2}\epsilon_{\parallel}(A_{1}-{2\over Z}\alpha
\chi_{1}^{z}-\alpha\chi_{1}\gamma_{\bf k})(\epsilon_{\parallel}
-\gamma_{\bf k})]\nonumber\\
&+&\lambda'^{2}[\alpha(\chi_{2}^{z}\gamma'_{\bf k}-{Z-1\over 2Z}
\chi_{2})\gamma'_{\bf k}+{1 \over 2}(A_{3}-{2 \over Z}\alpha
\chi_{2}^{z})]\nonumber\\
&+&\lambda\lambda'\alpha[\chi_{1}^{z}(1-
\epsilon_{\parallel}\gamma_{\bf k})\gamma'_{\bf k}+{1\over
2}(\chi_{1}\gamma'_{\bf k}-C_{2})(\epsilon_{\parallel} -\gamma_{\bf k})
\nonumber\\
&+&\gamma'_{\bf k}(C_{2}^{z}- \epsilon_{\parallel}
\chi_{2}^{z}\gamma_{\bf k})-{1 \over 2}\epsilon_{\parallel}
(C_{2}-\chi_{2}\gamma_{\bf k})] \nonumber\\
&+&\lambda J_{\rm eff \perp}\alpha[{1 \over 2}C_{\perp}
\epsilon_{\perp}({\bf k})(\epsilon_{\parallel}-\gamma_{\bf k})+{1
\over 2}\epsilon_{\parallel}\epsilon_{\perp}({\bf k})(C_{\perp}-
\chi_{\perp}\gamma_{\bf k})\nonumber\\
&+&C_{\perp}^{z}(1-\epsilon_{\parallel} \gamma_{\bf
k})+(C_{\perp}^{z}-\epsilon_{\parallel}\chi_{\perp}^{z} \gamma_{\bf
k})\nonumber\\
&+&(-1)^{\nu}({1 \over 2}\chi_{1}\epsilon_{\perp} ({\bf
k})(\epsilon_{\parallel}-\gamma_{\bf k})+{1 \over 2}
\epsilon_{\parallel}(C_{\perp}-\chi_{\perp}\gamma_{\bf k})\nonumber\\
&+&\chi_{1}^{z}\epsilon_{\perp}({\bf k})(1-\epsilon_{\parallel}
\gamma_{\bf k})+\epsilon_{\perp}({\bf
k})(C_{\perp}^{z}-\epsilon_{\parallel} \chi_{\perp}^{z}\gamma_{\bf
k}))]\nonumber\\
&+&\lambda' J_{\rm eff \perp} \alpha[\epsilon_{\perp}({\bf k})({1
\over 2}\chi_{\perp} \gamma'_{\bf
k}-C'_{\perp})+(C'^{z}_{\perp}+\chi_{\perp}^{z})
\gamma'_{\bf k}\nonumber\\
&+&(-1)^{\nu}(\epsilon_{\perp}({\bf k})(\chi_{2}^{z}+
\chi_{\perp}^{z})\gamma'_{\bf k}-{1 \over 2}(\chi_{2}
\epsilon_{\perp}({\bf k})-\chi_{\perp}\gamma'_{\bf
k})\nonumber\\
&-&{1\over 2} C'_{\perp})]+{1\over 4}J_{\rm eff \perp}^{2}
[\epsilon_{\perp}({\bf k})+ (-1)^{\nu}]^{2},
\end{eqnarray}
where $A_{1}=\alpha C_{1}+(1-\alpha)/2Z$, $A_{2}=\alpha C_{1}^{z}+
(1-\alpha)/4Z$, $A_{3}=\alpha C_{3}+(1-\alpha)/2Z$, and the spin
correlation functions $C_{1}=(1/Z^{2})
\sum_{\hat{\eta}\hat{\eta'}}\langle S_{i+\hat{\eta}a}^{+}
S_{i+\hat{\eta'}a}^{-}\rangle$, $C_{2}=(1/Z^{2})
\sum_{\hat{\eta}\hat{\tau}}\langle S_{i+\hat{\eta}a}^{+}
S_{i+\hat{\tau}a}^{-}\rangle$, $C_{3}=(1/Z^{2})
\sum_{\hat{\tau}\hat{\tau'}}\langle S_{i+\hat{\tau}a}^{+}
S_{i+\hat{\tau'}a}^{-}\rangle$, $C_{1}^{z}=(1/Z^{2})
\sum_{\hat{\eta}\hat{\eta'}}\langle S_{i+\hat{\eta}a}^{z}
S_{i+\hat{\eta'}a}^{z}\rangle$, $C_{2}^{z}=(1/Z^{2})
\sum_{\hat{\eta}\hat{\tau}}\langle S_{i+\hat{\eta}a}^{z}
S_{i+\hat{\tau}a}^{z}\rangle$, $C_{\perp}=(1/Z)\sum_{\hat{\eta}}
\langle S_{i1}^{+} S_{i+\hat{\eta}2}^{-}\rangle$,
$C{'}_{\perp}=(1/Z)\sum_{\hat{\tau}} \langle S_{i1}^{+}
S_{i+\hat{\tau}2}^{-}\rangle$, $C_{\perp}^{z}=(1/Z)
\sum_{\hat{\eta}}\langle S_{i1}^{z}S_{i+ \hat{\eta}2}^{z}\rangle$,
and $C{'}_{\perp}^{z}=(1/Z) \sum_{\hat{\tau}}\langle S_{i1}^{z}S_{i+
\hat{\tau}2}^{z}\rangle$. In order not to violate the sum rule of
the correlation function $\langle S^{+}_{i}S^{-}_{i}\rangle=1/2$ in
the case without AFLRO, the important decoupling parameter $\alpha$
has been introduced in the above MF calculation, which can be
regarded as the vertex correction \cite{kondo}.

For discussions of the electronic structure of doped bilayer
cuprates in the normal state, we need to calculate the electron
Green's function $G(i-j,t-t')=\langle\langle C_{i\sigma}(t);
C^{\dagger}_{j\sigma} (t')\rangle\rangle=G_{L}(i-j,t-t')
+\sigma_{x}G_{T}(i-j,t-t')$, which is a convolution of the spin
Green's function and dressed holon Green's function in the CSS
fermion-spin theory, and reflect the charge-spin recombination
\cite{anderson1}. In the present bilayer system, the longitudinal
and transverse parts of the electron Green's function can be
obtained explicitly in terms of the corresponding MF spin Green's
functions (7) and full dressed holon Green's functions (4) as,
\begin{subequations}
\begin{eqnarray}
&&G_{L}({\bf k},\omega)={1\over 8N}\sum_{\bf p}\sum_{\mu\nu}
Z_{FA}^{(\mu)}{B_{\nu{\bf p}}\over \omega_{\nu{\bf p}}}\nonumber\\
&&\times\left ({L^{(1)}_{\mu\nu}({\bf
k,p})\over\omega+\bar{\xi}_{\mu{\bf p-k}} -\omega_{\nu{\bf
p}}}+{L^{(2)}_{\mu\nu}({\bf k,p})\over\omega+
\bar{\xi}_{\mu{\bf p-k}}+\omega_{\nu{\bf p}}}\right ),\\
&&G_{T}({\bf k},\omega)={1\over 8N}\sum_{\bf p}\sum_{\mu\nu}
(-1)^{\mu+\nu}Z_{FA}^{(\mu)}{B_{\nu{\bf p}}\over\omega_{\nu{\bf p}
}}\nonumber\\
&&\times\left ({L^{(1)}_{\mu\nu}({\bf k,p})\over\omega+
\bar{\xi}_{\mu{\bf p-k}}-\omega_{\nu{\bf p}}}+
{L^{(2)}_{\mu\nu}({\bf k,p})\over\omega+ \bar{\xi}_{\mu{\bf p-k}}+
\omega_{\nu{\bf p}}}\right ),
\end{eqnarray}
\end{subequations}
where $L^{(1)}_{\mu\nu}({\bf k,p})=n_{F}(\bar{\xi}_{\mu{\bf p-k}})
+n_{B}(\omega_{\nu{\bf p}})$, $L^{(2)}_{\mu\nu}({\bf k,p})=1-n_{F}
(\bar{\xi}_{\mu{\bf p-k}})+n_{B}(\omega_{\nu{\bf p}})$, and
$n_{B}(\omega)$ and $n_{F}(\omega)$ are the boson and fermion
distribution functions, respectively, then the longitudinal and
transverse electron spectral functions $A_{L}({\bf k},\omega)=-
2{\rm Im}G_{L}({\bf k},\omega)$ and  $A_{T}({\bf k}, \omega) =-
2{\rm Im}G_{T}({\bf k},\omega)$ are obtained from the above
longitudinal and transverse electron Green's functions as,
\begin{subequations}
\begin{eqnarray}
A_{L}({\bf k},\omega)&=&\pi{1\over 4N}\sum_{\bf p}\sum_{\mu\nu}
Z_{FA}^{(\mu)}{B_{\nu{\bf p}}\over\omega_{\nu{\bf p}}}\nonumber\\
&\times&[L^{(1)}_{\mu\nu}({\bf k,p})\delta(\omega+\bar{\xi}_{\mu{\bf
p-k}} -\omega_{\nu{\bf p}})\nonumber\\
&&+L^{(2)}_{\mu\nu}({\bf k,p})\delta(\omega+
\bar{\xi}_{\mu{\bf p-k}}+\omega_{\nu{\bf p}})], \\
A_{T}({\bf k},\omega)&=&\pi{1\over 4N}\sum_{\bf p}\sum_{\mu\nu}
(-1)^{\mu+\nu}Z_{FA}^{(\mu)}{B_{\nu{\bf p}}\over\omega_{\nu{\bf p}}}
\nonumber\\
&\times&[L^{(1)}_{\mu\nu}({\bf k,p})\delta(\omega+\bar{\xi}_{\mu{\bf
p-k}}-\omega_{\nu{\bf p}})\nonumber\\
&&+L^{(2)}_{\mu\nu}({\bf k,p})\delta(\omega+ \bar{\xi}_{\mu{\bf
p-k}}+\omega_{\nu{\bf p}})].
\end{eqnarray}
\end{subequations}
With the help of the above longitudinal and transverse electron
spectral functions, the bonding and antibonding electron spectral
functions of doped bilayer cuprates are obtained as,
\begin{subequations}
\begin{eqnarray}
A^{+}({\bf k},\omega)&=&{1\over 2}[A_{L}({\bf k},\omega)+A_{T}
({\bf k},\omega)],\\
A^{-}({\bf k},\omega)&=&{1\over 2}[A_{L}({\bf k},\omega)-A_{T} ({\bf
k},\omega)],
\end{eqnarray}
\end{subequations}
respectively.

\section{Electron spectrum and quasiparticle dispersion of doped
bilayer cuprates}

We are now ready to discuss the doping and temperature dependence of
the electron spectrum and quasiparticle dispersion of bilayer
cuprate superconductors in the normal state. We have performed a
calculation for the electron spectral functions in Eq. (11), and the
results of the bonding (solid line) and antibonding (dashed line)
electron spectral functions at (a) the $[\pi,0]$ point and (b)
$[\pi/2,\pi/2]$ point for parameters $t/J=2.5$, $t'/t=0.15$, and
$t_{\perp}/t=0.3$ with temperature $T=0.1J$ at the doping
concentration $\delta=0.15$ are plotted in Fig. 1. Apparently, there
is a double-peak structure in the electron spectral function around
the $[\pi,0]$ point, i.e., the bonding and antibonding quasiparticle
peaks around the $[\pi,0]$ point are located at the different
positions, while the bonding and antibonding peaks around the
$[\pi/2,\pi/2]$ point are located at the same position, which leads
to that the BS appears around the $[\pi,0]$ point, and is absent
from the vicinity of the $[\pi/2,\pi/2]$ point. In particular, the
positions of the antibonding peaks at the $[\pi,0]$ point are more
closer to the Fermi energy than these for the bonding peaks. In this
sense, the differentiation between the bonding and antibonding
components of the electron spectral function around the $[\pi,0]$
point is essential. In analogy to the single layer cuprates
\cite{guo}, both positions of the quasiparticle peaks from the
bonding and antibonding electron spectral functions at the $[\pi,0]$
and $[\pi/2,\pi/2]$ points are below the Fermi energy, but the
positions of the peaks at the $[\pi/2,\pi/2]$ point are more closer
to the Fermi energy, which indicates that the lowest energy states
are located at the $[\pi/2,\pi/2]$ point, in other words, the low
energy spectral weight with the majority contribution to the
low-energy properties of doped bilayer cuprates comes from the
$[\pi/2,\pi/2]$ point, in qualitative agreement with the ARPES
experimental data on doped bilayer cuprates
\cite{shen,dfeng,kordyuk,chuang}. The double-peak structure in the
electron spectral functions around the $[\pi,0]$ point is closely
related to the interlayer hopping form in Eq. (2). With decreasing
the values of $t_{\perp}$ and $J_{\perp}$, the distance between the
bonding and antibonding peaks in the electron spectral functions
decreases. When $t_{\perp}=0$ and $J_{\perp}=0$, we find that the
transverse part of the dressed holon Green's functions in Eq. (4b)
(then the transverse part of the electron Green's functions in Eq.
(9b) and transverse part of the electron spectral functions in Eq.
(10b)) is equal to the zero. In this case, the bonding electron
spectral function in Eq. (11a) is exactly same as the antibonding
electron spectral function in Eq. (11b), then the electron spectral
functions are reduced to the doped single layer case \cite{guo}.

\begin{figure}
\includegraphics[scale=0.65]{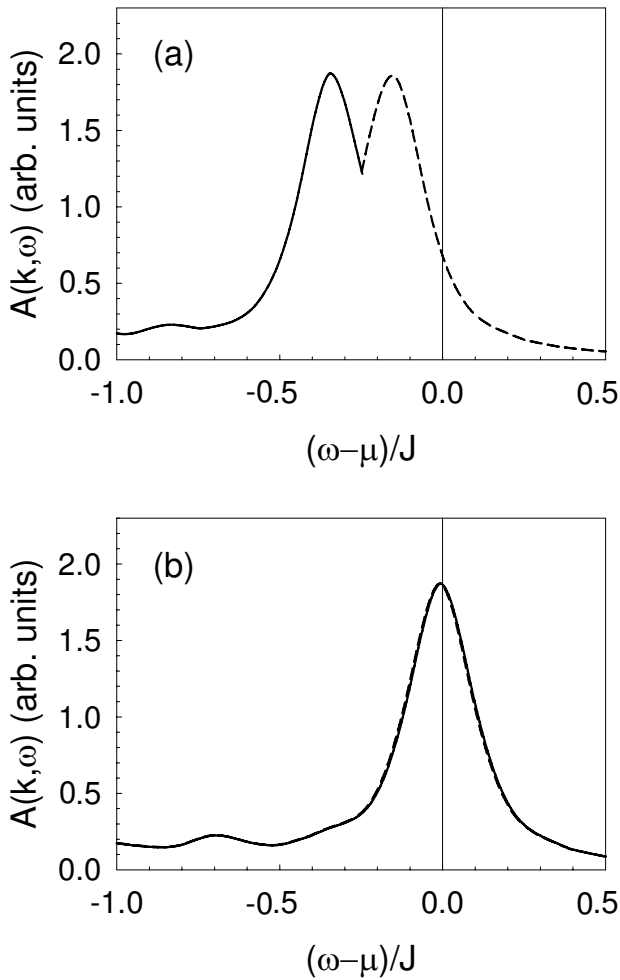}
\caption{The bonding (solid line) and antibonding (dashed line)
electron spectral functions as a function of energy at (a) the
$[\pi,0]$ point and (b) $[\pi/2,\pi/2]$ point for $t/J=2.5$,
$t'/t=0.15$, and $t_{\perp}/t=0.3$ with $T=0.1J$ at $\delta=0.15$.}
\end{figure}

For a better understanding of the physical properties of the
electron spectrum in doped bilayer cuprates, we have studied the
electron spectrum at different doping concentrations, and the result
of the electron spectral functions at $[\pi,0]$ point for $t/J=2.5$,
$t'/t=0.15$, and $t_{\perp}/t=0.3$ with $T=0.1J$ at $\delta=0.09$
(solid line), $\delta=0.12$ (dashed line), and $\delta=0.15$ (dotted
line) are plotted in Fig. 2, which indicates that with increasing
the doping concentration, both bonding and antibonding quasiparticle
peaks become sharper, and the spectral weights of these peaks
increase in intensity. Furthermore, we have also discussed the
temperature dependence of the electron spectrum, and the result of
the electron spectral functions at $[\pi,0]$ point for $t/J=2.5$,
$t'/t=0.15$, and $t_{\perp}/t=0.3$ at $\delta=0.15$ with $T=0.1J$
(solid line), $T=0.05J$ (dashed line), and $T=0.01J$ (dotted line)
are plotted in Fig. 3, it is shown obviously that both bonding and
antibonding spectral weights are suppressed with increasing
temperatures. Our these results are also qualitatively consistent
with the ARPES experimental results on doped bilayer cuprates
\cite{shen,dfeng,kordyuk,chuang}.

\begin{figure}
\includegraphics[scale=0.56]{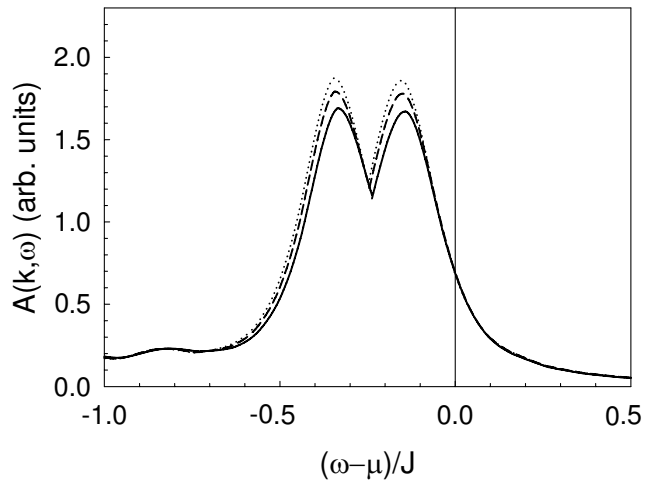}
\caption{The electron spectral functions at $[\pi,0]$ point for
$t/J=2.5$, $t'/t=0.15$, and $t_{\perp}/t=0.3$ with $T=0.1J$ at
$\delta=0.09$ (solid line), $\delta=0.12$ (dashed line), and
$\delta=0.15$ (dotted line).}
\end{figure}

\begin{figure}
\includegraphics[scale=0.56]{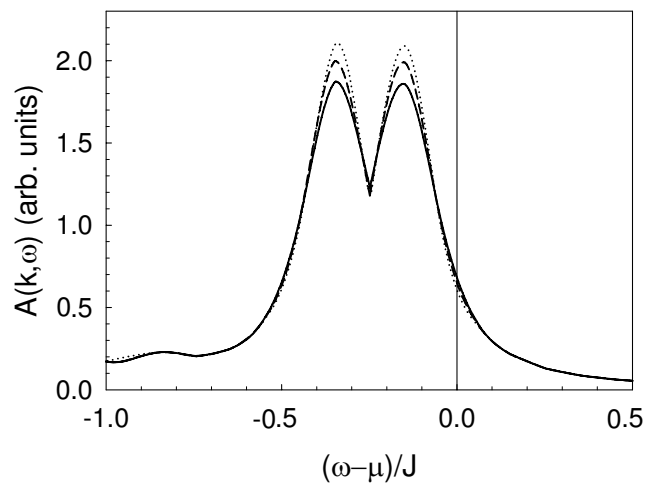}
\caption{The electron spectral functions at $[\pi,0]$ point for
$t/J=2.5$, $t'/t=0.15$, and $t_{\perp}/t=0.3$ at $\delta=0.15$ with
$T=0.1J$ (solid line), $T=0.05J$ (dashed line), and $T=0.01J$
(dotted line).}
\end{figure}

For considering the quasiparticle dispersion of doped bilayer
cuprates, we have made a series of calculations for both bonding and
antibonding electron spectral functions at different momenta, and
find that the lowest energy peaks are well defined at all momenta.
In particular, the positions of both bonding and antibonding
quasiparticle peaks as a function of energy $\omega$ for momentum
${\bf k}$ in the vicinity of the $[\pi,0]$ point are almost not
changeable, which leads to the unusual quasiparticle dispersion
around the $[\pi,0]$ point. To show this broad feature clearly, we
plot the positions of the lowest energy quasiparticle peaks in the
bonding and antibonding electron spectra as a function of momentum
along the high symmetry directions with $T=0.1J$ at $\delta=0.15$
for $t/J=2.5$, $t'/t=0.15$, and $t_{\perp}/t=0.3$ in Fig. 4. For
comparison, the corresponding result from the tight binding fit to
the experimental data of the doped bilayer cuprate
Bi$_{2}$Sr$_{2}$CaCu$_{2}$O$_{8+\delta}$ \cite{kordyuk} is also
shown in Fig. 2 (inset). Our result shows that in analogy to the
doped single layer cuprates \cite{guo}, both electron bonding and
antibonding quasiparticles around the $[\pi,0]$ point disperse very
weakly with momentum, and then the two main flat bands appear, while
the Fermi energy is only slightly above these flat bands. Moreover,
this bilayer energy band splitting reaches its maximum at the
$[\pi,0]$ point. Our this result shows that the bilayer interaction
has significant contributions to the electronic structure of doped
bilayer cuprates, and is in qualitative agreement with these
obtained from the ARPES experimental measurements on doped bilayer
cuprates \cite{shen,dfeng,kordyuk,chuang}. Within the
$t$-$t'$-$t''$-$J$ model, this quasiparticle dispersion of doped
bilayer cuprates has been studied based on the resonating valence
bond wave function and Gutzwiller approximation \cite{mori}, where
the quasiparticle dispersion relations are degenerate along the
$[0,0]$ to $[\pi,\pi]$ direction, and the splitting becomes maximum
at the $[\pi,0]$ point, which is consistent with our present result.

\begin{figure}
\includegraphics[scale=0.50]{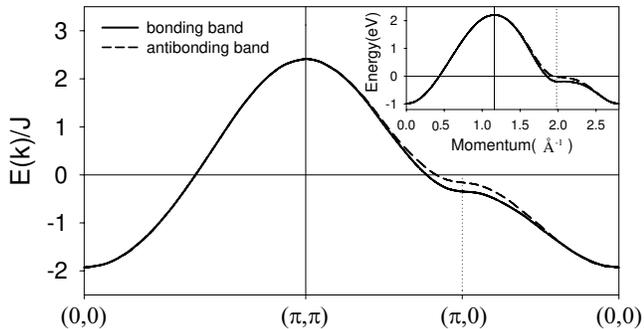}
\caption{The positions of the lowest energy quasiparticle peaks in
the bonding (solid line) and antibonding (dashed line) electron
spectra as a function of momentum with $T=0.1J$ at $\delta=0.15$ for
$t/J=2.5$, $t'/t=0.15$, and $t_{\perp}/t=0.3$. Inset: the
corresponding result from the tight binding fit to the experimental
data of the doped bilayer cuprate
Bi$_{2}$Sr$_{2}$CaCu$_{2}$O$_{8+\delta}$ taken from Ref.
\protect\onlinecite{kordyuk}.}
\end{figure}

The essential physics of the double-peak structure of the electron
spectral function around the $[\pi,0]$ point in the present doped
bilayer cuprates is dominated by the bilayer interaction. The full
electron Green's function in doped bilayer cuprates is divided into
the longitudinal and transverse parts, respectively, due to the
bilayer interaction, then these longitudinal and transverse Green's
function (then the bonding and antibonding electron spectral
functions and corresponding quasiparticle dispersions) are obtained
beyond the MF approximation by considering the dressed holon
fluctuation due to the spin pair bubble, therefore the nature of the
bonding and antibonding electron spectral functions of doped bilayer
cuprates in the normal state are closely related to the strong
interaction between the dressed holons (then electron
quasiparticles) and collective magnetic excitations. In this case,
the single-particle hoppings in the bilayer $t$-$t'$-$J$ model are
strongly renormalized by the magnetic interaction. As a consequence,
both bonding and antibonding quasiparticle bandwidths are reduced to
the order of (a few) $J$, and then the energy scales of both bonding
and antibonding quasiparticle energy bands are controlled by the
magnetic interaction. These renormalizations for both bonding and
antibonding energy bands are then responsible for the unusual
bonding and antibonding electron quasiparticle spectra and
production of the two main flat bands around the $[\pi,0]$ point. As
in the single layer case, our present results also show that the
electron quasiparticle excitations in doped bilayer cuprates
originating from the dressed holons and spins are due to the
charge-spin recombination, this reflects the composite nature of the
electron quasiparticle excitations, and then the unconventional
normal-state properties in doped bilayer cuprates are attributed to
the presence of the dressed holons, spins, electron quasiparticle
excitations, and bilayer coupling.

\section{Summary and discussions}

In conclusion we have shown that if the strong dressed holon-spin
coupling and bilayer interaction are taken into account, the
$t$-$t'$-$J$ model per se can correctly reproduce some main features
of the ARPES electron spectrum and quasiparticle dispersion of doped
bilayer cuprates in the normal state. In our opinion, the
differentiation between the bonding and antibonding components is
essential, which leads to two main flat bands around the $[\pi,0]$
point below the Fermi energy. In analogy to the doped single layer
cuprates, the lowest energy states in doped bilayer cuprates are
located at the $[\pi/2,\pi/2]$ point. Our results also show that the
striking behavior of the electronic structure in doped bilayer
cuprates is intriguingly related to the bilayer interaction together
with strong coupling between the electron quasiparticles and
collective magnetic excitations.

The experimental \cite{tanaka} and theoretical
\cite{chakarvarty,mori} analysis shows that the physical properties
for different families of doped cuprates are strongly correlated
with $t'$ and $t_{\perp}$, this means that the different values of
parameters $t/J$, $t'/t$, and $t_{\perp}/t$ should be chosen for
different families of doped cuprates. For comparison, the values of
parameters $t/J$ and $t'/t$ in the present work have been chosen as
the same as in the single layer case \cite{guo}. Furthermore, the
value of parameter $t_{\perp}/t$ in the present work is
qualitatively consistent with these in the local density
approximation calculations \cite{chakarvarty} based on the bilayer
Hubbard model and the calculations within the bilayer
$t$-$t'$-$t''$-$J$ model based on the resonating valence bond wave
function and Gutzwiller approximation \cite{mori}. Although the
simplest bilayer $t$-$t'$-$J$ model can not be regarded as a
complete model for the quantitative comparison with doped bilayer
cuprates, our present results for the normal state are in
qualitative agreement with the major experimental observations in
the normal state of doped bilayer cuprates
\cite{shen,dfeng,kordyuk,chuang}.

Finally, we have noted that the very recent ARPES experiments have
shown that although the largest BS is found in the electron spectral
function around the $[\pi,0]$ point, a small BS remains also around
the $[\pi/2,\pi/2]$ point in a wide doping range
\cite{kordyuk2,shen3}. It has been argued that the observed this
small BS around the $[\pi/2,\pi/2]$ point is in good agreement with
the local density approximation based band structure calculations,
and is caused by the vertical O $2p_{\sigma}$-O $2p_{\sigma}$
hopping $t^{pp}_{\perp}$ between two adjacent planes
\cite{kordyuk2,shen3}. Therefore the significance of the $p-p$
transfer $t^{pp}_{\perp}$ should be taken into account. These and
related issues are under investigations now.

\acknowledgments The authors would like to thank Dr. H. Guo and Dr.
L. Cheng for the helpful discussions. This work was supported by the
National Natural Science Foundation of China under Grant Nos.
90403005 and 10547104, and the funds from the Ministry of Science
and Technology of China under Grant Nos. 2006CB601002 and
2006CB921300.

\appendix

\section*{The dressed holon Green's function in the bilayer system}

As in the single layer case \cite{guo}, the full dressed holon
Green's function in the doped bilayer cuprates satisfies the
equation \cite{eliashberg},
\begin{eqnarray}
g({\bf k},\omega)=g^{(0)}({\bf k},\omega)+g^{(0)}({\bf k},\omega)
\Sigma^{(h)}({\bf k},\omega)g({\bf k},\omega)
\end{eqnarray}
with the MF dressed holon Green's function $g^{(0)}({\bf k},
\omega)=g^{(0)}_{L}({\bf k},\omega)+\sigma_{x}g^{(0)}_{T}({\bf k},
\omega)$, where the longitudinal and transverse parts are obtained
as,
\begin{subequations}
\begin{eqnarray}
g^{(0)}_{L}({\bf k},\omega)&=&{1\over 2}\sum_{\nu=1,2}{1\over
\omega-\xi_{\nu{\bf k}}},\\
g^{(0)}_{T}({\bf k},\omega)&=&{1\over 2}\sum_{\nu=1,2}(-1)^{\nu+1}
{1\over\omega-\xi_{\nu{\bf k}}},
\end{eqnarray}
\end{subequations}
and the dressed holon self-energy function $\Sigma^{(h)}({\bf k},
\omega)=\Sigma^{(h)}_{L}({\bf k},\omega)+\sigma_{x}\Sigma^{(h)}_{T}
({\bf k},\omega)$, where $\Sigma^{(h)}_{L}({\bf k},\omega)$ and
$\Sigma^{(h)}_{T}({\bf k},\omega)$ are the corresponding
longitudinal and transverse parts, and have been given in Eq. (6).
This self-energy function $\Sigma^{(h)}({\bf k},\omega)$
renormalizes the MF dressed holon spectrum, and thus it describes
the quasiparticle coherence. On the other hand, since we only study
the low-energy behavior of doped bilayer cuprates, then the
quasiparticle coherence can be discussed in the static limit. In
this case, we follow the previous discussions for the single layer
case \cite{guo}, and obtain explicitly the longitudinal and
transverse parts of the full dressed holon Green's functions of the
doped bilayer system as,
\begin{subequations}
\begin{eqnarray}
g_{L}({\bf k},\omega)&=&{1\over 2}\sum_{\nu=1,2}{Z^{(\nu)}_{FA}
\over\omega-\bar{\xi}_{\nu{\bf k}}}, \\
g_{T}({\bf k},\omega)&=&{1\over 2}\sum_{\nu=1,2}(-1)^{\nu+1}{
Z^{(\nu)}_{FA }\over \omega-\bar{\xi}_{\nu{\bf k}}},
\end{eqnarray}
\end{subequations}
with the quasiparticle coherent weights satisfy the following
equations,
\begin{subequations}
\begin{eqnarray}
{1\over Z_{FA}^{(1)}}&=&1+{1\over 32N^{2}}\sum_{{\bf q,p}}
\sum_{\nu,\nu',\nu''}[1+(-1)^{\nu+\nu'+\nu''+1}]\nonumber\\
&\times&\Lambda_{\nu\nu'\nu''}({\bf q,p}),\\
{1\over Z_{FA}^{(2)}}&=&1+{1\over 32N^{2}}\sum_{{\bf q,p}}
\sum_{\nu,\nu',\nu''}[1-(-1)^{\nu+\nu'+\nu''+1}]\nonumber\\
&\times&\Lambda_{\nu\nu'\nu''}({\bf q,p}),
\end{eqnarray}
\end{subequations}
where the kernel function $\Lambda_{\nu\nu'\nu''}({\bf q,p})$ can be
evaluated as,
\begin{eqnarray}
\Lambda_{\nu\nu'\nu''}({\bf q,p})&=&C_{\nu\nu''}({\bf p+k_{0}})
Z^{(\nu'')}_{FA}{B_{\nu'{\bf p}}B_{\nu{\bf q}}\over \omega_{\nu'{\bf
p}} \omega_{\nu{\bf q}}}\nonumber\\
&\times&\left( {F^{(1)}_{\nu\nu'\nu''}({\bf q,p})\over
[\omega_{\nu'{\bf p} } -\omega_{\nu{\bf q}}+\bar{\xi}_{\nu''{\bf
p-q+k_{0}}}]^{2}}\right.\nonumber\\
&&\left.+{F^{(2)}_{\nu\nu'\nu''}({\bf q,p})\over [\omega_{\nu'{\bf
p}}- \omega_{\nu{\bf q}}-\bar{\xi}_{\nu''{\bf
p-q+k_{0}}}]^{2}}\right.\nonumber\\
&&\left. +{F^{(3)}_{\nu\nu'\nu''}({\bf q,p})\over [\omega_{\nu'{\bf
p}} +\omega_{\nu{\bf q}}+\bar{\xi}_{\nu''{\bf p-q+k_{0}}}]^{2}}\right.
\nonumber\\
&&\left.+{F^{(4)}_{\nu\nu'\nu''}({\bf q,p})\over [\omega_{\nu'{\bf
p}}+\omega_{\nu{\bf q}}-\bar{\xi}_{\nu''{\bf p- q+k_{0}}}]^{2}}
\right),~~~~
\end{eqnarray}
with $C_{\nu\nu''}({\bf p})=[Z(t\gamma_{\bf p}-t'\gamma{'}_{\bf p}
)+(-1)^{\nu+\nu''}t_{\perp}({\bf p})]^{2}$,
$F^{(1)}_{\nu\nu'\nu''}({\bf q,p})=n_{F}(\bar{\xi}_{\nu''{\bf p-
q+k_{0}}}) [n_{B}(\omega_{\nu'{\bf p}})-n_{B}(\omega_{\nu{\bf q}})
]+n_{B}(\omega_{\nu{\bf q}})[1+n_{B}(\omega_{\nu'{\bf p}})]$,
$F^{(2)}_{\nu\nu'\nu''}({\bf q,p})=n_{F}(\bar{\xi}_{\nu''{\bf p-
q+k_{0}}}) [n_{B}(\omega_{\nu{\bf q}})-n_{B}(\omega_{\nu'{\bf p}})
]+ n_{B}(\omega_{\nu'{\bf p}})[1+n_{B}(\omega_{\nu{\bf q}})]$,
$F^{(3)}_{\nu\nu'\nu''}({\bf q,p})=[1-n_{F}(\bar{\xi}_{\nu''{\bf p
-q+k_{0}}})][1+n_{B}(\omega_{\nu{\bf q}})+n_{B}(\omega_{\nu'{\bf p
}})]+n_{B}(\omega_{\nu{\bf q}})n_{B}(\omega_{\nu'{\bf p}})$,
$F^{(4)}_{\nu\nu'\nu''}({\bf q,p})=n_{F}(\bar{\xi}_{\nu''{\bf p-
q+k_{0}}}) [1+n_{B}(\omega_{\nu{\bf q}})+n_{B}(\omega_{\nu'{\bf p}
})]+ n_{B}(\omega_{\nu{\bf q}})n_{B}(\omega_{\nu'{\bf p}})$. These
two equations in Eq. (A4) must be solved together with other
self-consistent equations \cite{guo},
\begin{subequations}
\begin{eqnarray}
\delta&=&{1\over 4N}\sum_{\nu,{\bf k}}Z^{(\nu)}_{FA}\left(1-{\rm th}
[{1\over 2}\beta \bar{\xi}_{\nu{\bf k}}]\right ), \\
\phi_{1}&=&{1\over 4N}\sum_{\nu,{\bf k}}\gamma_{\bf k}Z^{(\nu)}_{FA}
\left ( 1-{\rm th}[{1\over 2}\beta \bar{\xi}_{\nu{\bf k}}]\right ),\\
\phi_{2}&=&{1\over 4N}\sum_{\nu,{\bf k}}\gamma{'}_{\bf k}
Z^{(\nu)}_{FA}\left ( 1-{\rm th}[{1\over 2}\beta
\bar{\xi}_{\nu{\bf k}}]\right ), \\
\phi_{\perp}&=&{1\over 4N}\sum_{\nu,{\bf k}} (-1)^{\nu+1}
Z^{(\nu)}_{FA} \left ( 1-{\rm th}[{1\over 2}\beta
\bar{\xi}_{\nu{\bf k}}]\right ), ~~~~~\\
1\over 2&=&{1\over 4N}\sum_{\nu,{\bf k}}{B_{\nu{\bf k}}\over
\omega_{\nu{\bf k}}}{\rm coth}[{1\over 2}\beta\omega_{\nu{\bf k}}],\\
\chi_{1} &=& {1\over 4N} \sum_{\nu,{\bf k}}\gamma_{\bf k}
{B_{\nu{\bf k}}\over\omega_{\nu{\bf k}}}{\rm coth}[{1\over 2}
\beta\omega_{\nu{\bf k}}], \\
\chi_{2}&=&{1\over 4N}\sum_{\nu,{\bf k}} \gamma{'}_{\bf k}
{B_{\nu{\bf k}}\over\omega_{\nu{\bf k}}}{\rm coth}
[{1\over 2}\beta\omega_{\nu{\bf k}}], \\
C_{1} &=& {1\over 4N} \sum_{\nu,{\bf k}} \gamma_{\bf k}^{2}
{B_{\nu{\bf k}}\over\omega_{\nu{\bf k}}}{\rm coth}
[{1\over 2}\beta\omega_{\nu{\bf k}}], \\
C_{2}&=&{1\over 4N}\sum_{\nu,{\bf k}}\gamma_{\bf k}\gamma{'}_{\bf k}
{B_{\nu{\bf k}}\over\omega_{\nu{\bf k}}}{\rm coth}
[{1\over 2}\beta\omega_{\nu{\bf k}}], \\
C_{3}&=&{1\over 4N}\sum_{\nu,{\bf k}} \gamma{'}_{\bf k}^{2}
{B_{\nu{\bf k}}\over\omega_{\nu{\bf k}}}{\rm coth}
[{1\over 2}\beta\omega_{\nu{\bf k}}], \\
\chi_{1}^{z}&=&{1\over 4N}\sum_{\nu,{\bf k}} \gamma_{\bf k}
{B_{z\nu{\bf k}}\over\omega_{z\nu{\bf k}}}{\rm coth}
[{1\over 2}\beta\omega_{z\nu{\bf k}}], \\
\chi_{2}^{z}&=&{1\over 4N}\sum_{\nu,{\bf k}} \gamma{'}_{\bf k}
{B_{z\nu{\bf k}}\over\omega_{z\nu{\bf k}}}{\rm coth}
[{1\over 2}\beta\omega_{z\nu{\bf k}}], \\
C_{1}^{z}&=&{1\over 4N}\sum_{\nu,{\bf k}} \gamma_{\bf k}^{2}
{B_{z\nu{\bf k}}\over\omega_{z\nu{\bf k}}}{\rm coth}
[{1\over 2}\beta\omega_{z\nu{\bf k}}], \\
C_{2}^{z}&=&{1\over 4N}\sum_{\nu,{\bf k}}\gamma_{\bf k}
\gamma{'}_{\bf k}   {B_{z\nu{\bf k}} \over \omega_{z\nu{\bf k}}}
{\rm coth}[{1\over 2}\beta\omega_{z\nu{\bf k}}], \\
\chi_{\perp} &=& {1\over 4N}\sum_{\nu,{\bf k}}(-1)^{\nu+1}
{B_{\nu{\bf k}}\over\omega_{\nu{\bf k}}}{\rm coth}[{1\over
2}\beta\omega_{\nu{\bf k}}], \\
C_{\perp} &=& {1\over 4N} \sum_{\nu,{\bf k}} (-1)^{\nu+1}
\gamma_{\bf k}{B_{\nu{\bf k}}\over\omega_{\nu{\bf k}}}
{\rm coth}[{1\over 2}\beta\omega_{\nu{\bf k}}], \\
C{'}_{\perp}&=&{1\over 4N}\sum_{\nu,{\bf k}}(-1)^{\nu+1}
\gamma{'}_{\bf k}{B_{\nu{\bf k}}\over\omega_{\nu{\bf k}}}
{\rm coth}[{1\over 2}\beta\omega_{\nu{\bf k}}], \\
\chi^{z}_{\perp}&=&{1\over 4N}\sum_{\nu,{\bf k}}(-1)^{\nu+1}
{B_{z\nu{\bf k}}\over\omega_{z\nu{\bf k}}}{\rm coth}[{1\over
2}\beta\omega_{z\nu{\bf k}}],\\
C_{\perp}^{z}&=&{1\over 4N}\sum_{\nu,{\bf k}}(-1)^{\nu+1}
\gamma_{\bf k} {B_{z\nu{\bf k}} \over \omega_{z\nu{\bf k}}}
{\rm coth}[{1\over 2}\beta\omega_{z\nu{\bf k}}],\\
C{'}_{\perp}^{z}&=&{1\over 4N}\sum_{\nu,{\bf k}}(-1)^{\nu+1}
\gamma{'}_{\bf k}  {B_{z\nu{\bf k}}  \over \omega_{z\nu{\bf k}}}
{\rm coth}[{1\over 2}\beta\omega_{z\nu{\bf k}}],
\end{eqnarray}
\end{subequations}
then all the order parameters, decoupling parameter $\alpha$,
chemical potential $\mu$ and the quasiparticle coherent weights are
obtained by the self-consistent calculation.

{}

\end{document}